\documentclass[aps,prl,twocolumn,groupedaddress]{revtex4}
\usepackage{graphicx}
\usepackage{color}

\def\beq{\begin{equation}}
\def\eeq{\end{equation}}
\def\beqr{\begin{eqnarray}}
\def\eeqr{\end{eqnarray}}
\def\bdpm{\begin{displaymath}}
\def\edpm{\end{displaymath}}

\begin{document}


\title{Model independent analysis of the forward-backward asymmetry of \\ 
top quark production at the Tevatron}



\author{Dong-Won Jung}
\affiliation{Physics Department and CMTP, National Central University, 
Jhongli, Taiwan, 32054}

\author{P. Ko}
\affiliation{School of Physics, KIAS, Seoul 130-722, Korea}

\author{Jae Sik Lee}
\affiliation{Physics Division, National Center for Theoretical Sciences, 
Hsinchu, Taiwan 300}

\author{Soo-hyeon Nam}
\affiliation{Korea Institute of Science and Technology Information, 
Daejeon 305-806, Korea}

\date{December 8, 2009}

\begin{abstract}
Motivated by a possible anomaly in the forward-backward (FB) asymmetry 
of top quark ($A_{\rm FB}$) observed at the Tevatron, we perform a model 
independent analysis on $q \bar{q} \rightarrow t \bar{t}$ using an effective 
lagrangian with dim-6 four-quark operators.  
We derive necessary conditions on new physics structures 
and the couplings that are consistent with the $t\bar{t}$ production 
cross section and $A_{\rm FB}$ measured at the Tevatron, and 
discuss possible new physics scenarios that could generate such
dim-6 operators.
\end{abstract}


\maketitle



{\bf\large 1.} 
Top physics has entered a new era after its first discovery, 
due to the high luminosity achieved at the Tevatron.  
Most recent results on the top mass and the $t\bar{t}$ production 
cross section (CDF and D0 Collaborations combined analysis) are:
$m_t  =   ( 171.3 \pm 1.3 ) ~{\rm GeV}$  and 
$\sigma_{t\bar{t}}  =   ( 7.50 \pm 0.48 ) ~{\rm pb}$, respectively  
\cite{cdf2009}.

The forward-backward asymmetry $A_{\rm FB}$ of the top quark is 
one of the interesting observables related with top quark. 
Within the Standard Model (SM), this asymmetry vanishes at leading order 
in  QCD because of $C$ symmetry. 
At next-to-leading order [$O(\alpha_s^3)$], 
a nonzero $A_{\rm FB}$  can develop from the interference
between the Born amplitude and two-gluon intermediate state, 
as well as the gluon bremsstrahlung and gluon-(anti)quark scattering 
into $t \bar{t}$, 
with the prediction $A_{\rm FB}\sim 0.078$ \cite{Antunano:2007da}.  
The measured asymmetry has been off the SM prediction by $2 \sigma$ 
for the last few years, albeit a large experimental uncertainties. 
The most recent measurement in the $t\bar{t}$ rest frame is 
\cite{cdf2009}
\begin{equation}
A_{\rm FB} \equiv \frac{N_t ( \cos\theta \geq 0) - N_{\bar{t}} 
( \cos\theta \geq 0 )}{N_t ( \cos\theta \geq 0) + N_{\bar{t}} 
( \cos\theta \geq 0 )} =  (0.24 \pm 0.13 \pm 0.04) 
\end{equation}
with $\theta$ being the polar angle of the top quark with 
respect to the incoming proton in the $t\bar{t}$ rest frame.
This $\sim 2\sigma$ deviation stimulated some speculations on new physics
scenarios \cite{Choudhury:2007ux,Djouadi:2009nb,Ferrario:2009bz,
Jung:2009jz,Cheung:2009ch,Frampton:2009rk,Shu:2009xf}. 

On the other hand, search for a new resonance decaying into  
$t\bar{t}$ pair has been  carried out  at the Tevatron. 
As of now, there is no clear signal for such a new resonance \cite{cdf2009}.  
Therefore, in this letter, we assume that  a new physics scale relevant to 
$A_{\rm FB}$ is large enough so that  production of a new particle is 
beyond the reach of the Tevatron, which makes a key difference between 
our work and other literatures on this subject 
\cite{Choudhury:2007ux,Djouadi:2009nb,Ferrario:2009bz,Jung:2009jz,
Cheung:2009ch,Frampton:2009rk,Shu:2009xf}.  
Then it is adequate to integrate out the heavy fields, and we can adopt 
a model independent effective lagrangian approach in order to study 
new physics effects on $\sigma_{t\bar{t}}$ and $A_{\rm FB}$.  
The resulting effective lagrangian will be an infinite tower of local 
operators containing light quark ($q=u,d,s,c,b$), (anti)top quark 
($t$ and $\overline{t}$) and gluon fields.  
At the Tevatron, the $t\bar{t}$ production is dominated by $q\bar{q} 
\rightarrow t\bar{t}$, and it would sufficient to consider dimension-6
four-quark operators (the so-called contact interaction terms) 
to describe the new physics effects on the $t\bar{t}$ production 
at the Tevatron.  Note that similar approach was
adopted for the dijet production to constrain the composite scale of 
light quarks, and we are proposing the same analysis for the $t\bar{t}$ 
system.  We will also speculate underlying new physics which could induce 
the relevant dim-6 operators in the second half of this work.

\medskip
{\bf\large 2.} If new physics scale is high enough, then their effects on the $t\bar{t}$ 
production at the Tevatron can be described by dim-6 effective lagrangian. 
Since the $SU(3)_C \times SU(2)_L \times U(1)_Y $
symmetry has been well established for the light quark system,  
we assume that $SU(2)_L \times U(1)_Y$ symmetry is linearly realized
on the light quark system.  And we impose the custodial symmetry 
$SU(2)_R$ for the light quark sector.  
Under these assumptions, the dimension-6 operators relevant to the 
$t\bar{t}$ production at the Tevatron are
\begin{eqnarray}
\mathcal{L}_6 &=& \frac{g_s^2}{\Lambda^2}\sum_{A,B}
\left[C^{AB}_{1q}(\bar{q}_A\gamma_\mu  q_A)(\bar{t}_B\gamma^\mu t_B) \right.
\nonumber \\
&& \hspace{1.0cm} + \left.
C^{AB}_{8q}(\bar{q}_A T^a\gamma_\mu q_A)(\bar{t}_B T^a\gamma^\mu  t_B)\right]
\end{eqnarray}
where $T^a = \lambda^a /2$, $\{A,B\}=\{L,R\}$, and 
$L,R \equiv (1 \mp \gamma_5)/2$ 
with $q=(u,d)^T,(c,s)^T$ 
\footnote{Although we assume the $SU(2)_L \times SU(2)_R$ chiral symmetry
for light quarks, all the explicit models do not satisfy this condition. 
In that case, one can interpret $q=u,d,s,c,b$.}.
Using this effective lagrangian, we calculate the cross section up to 
$O(1/\Lambda^2)$, keeping only the interference term between 
the SM and new physics contributions.  
In this case, we can safely neglect the flavor changing dim-6 
operators such as 
$\overline{d_R} \gamma^\mu s_R \overline{t_R} \gamma_\mu t_R$,
because they do not interfere with the SM QCD amplitude. 

The above effective lagrangian was also discussed in 
Ref.~\cite{Hill:1993hs}, where the $t$ quark was treated as 
$SU(2)_L \times SU(2)_R$ singlet and top currents were 
decomposed into vector and axial vector currents, rather than
chirality basis as in our case.   
If we include $t_L$ in (2),  we better work in terms of 
$Q_L^3 \equiv  ( t_L , b_L )^T$, since they forms an $SU(2)_L$ doublet.
Then we have to consider $b\bar{b}$ too. Note that the $b\bar{b}$ 
cross section agrees well with the SM predictions, and there are no 
data on the FB asymmetry of bottom quark at hadron colliders. 
We can avoid the complication from $b\bar{b}$  system,  
assuming that only $t_R$ appears in the above effective lagrangian.   
In fact, there are a number of new physics scenarios, where 
$t_R$ (and/or Higgs) is (partially) composite \cite{Georgi:1994ha}.
Then among the dim-6 four-quark operators, those who have 
$t_R$ and $\bar{t}_R$ would be most important at low energy.  
The case of composite $t_R$ can be addressed by choosing 
$C_{1q}^{LL} = C_{1q}^{RL} = C_{8q}^{LL} = C_{8q}^{RL}=0$

\begin{figure*}
\includegraphics[width=15cm]{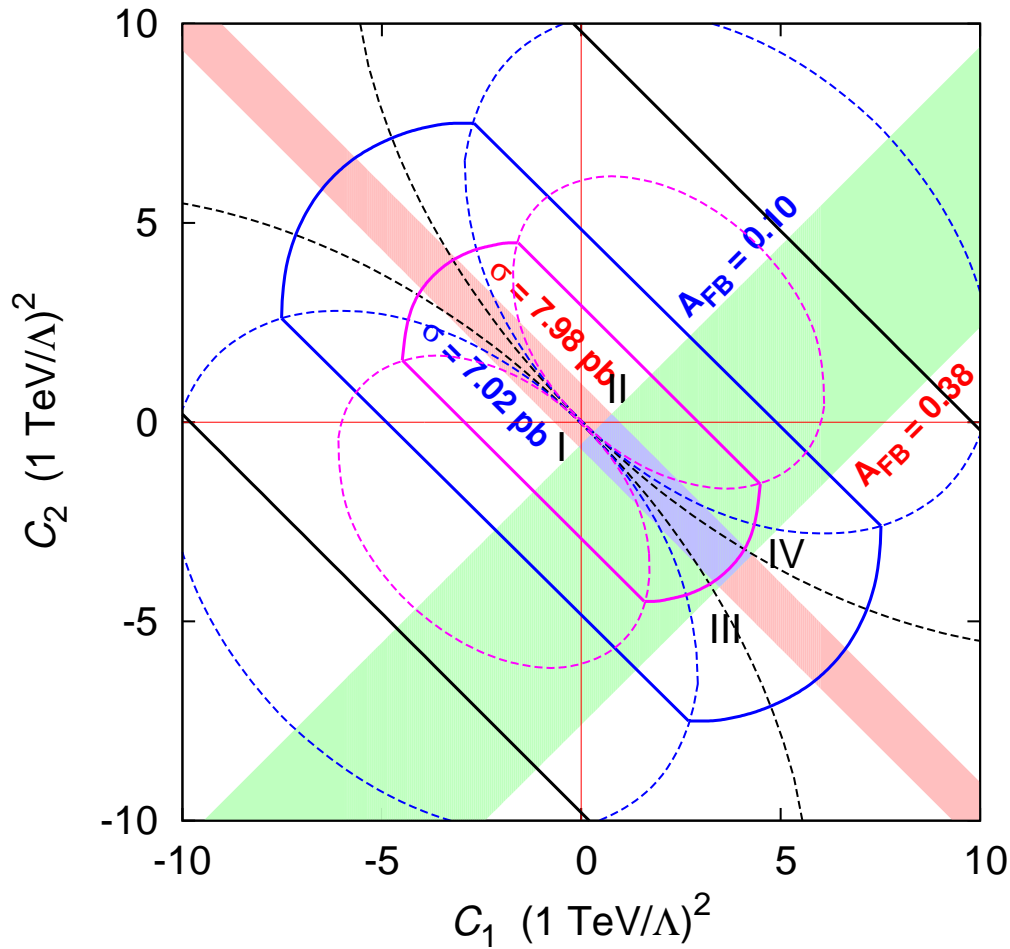}\\%
\caption{\label{} The region in $( C_1 , C_2 )$ plane that is
consistent with the Tevatron data  at the $1 \sigma$ level:   
$\sigma_{t\bar{t}} = (7.50 \pm 0.48)$ pb  
and $A_{\rm FB} = (0.24 \pm 0.13 \pm 0.04)$.
Also shown are the boundaries of the validity regions described in the text taking
$r=0.3$ (magenta), $0.5$ (blue) and $1$ (black). 
The straight lines are from the inequality Eq.~(4), 
the circles and their mirror images from the inequality Eq.~(5), and 
the ellipses centered at the origin from Eq.~(6).
}
\label{fig:1a}
\end{figure*}
\begin{figure}
\includegraphics[width=7cm]{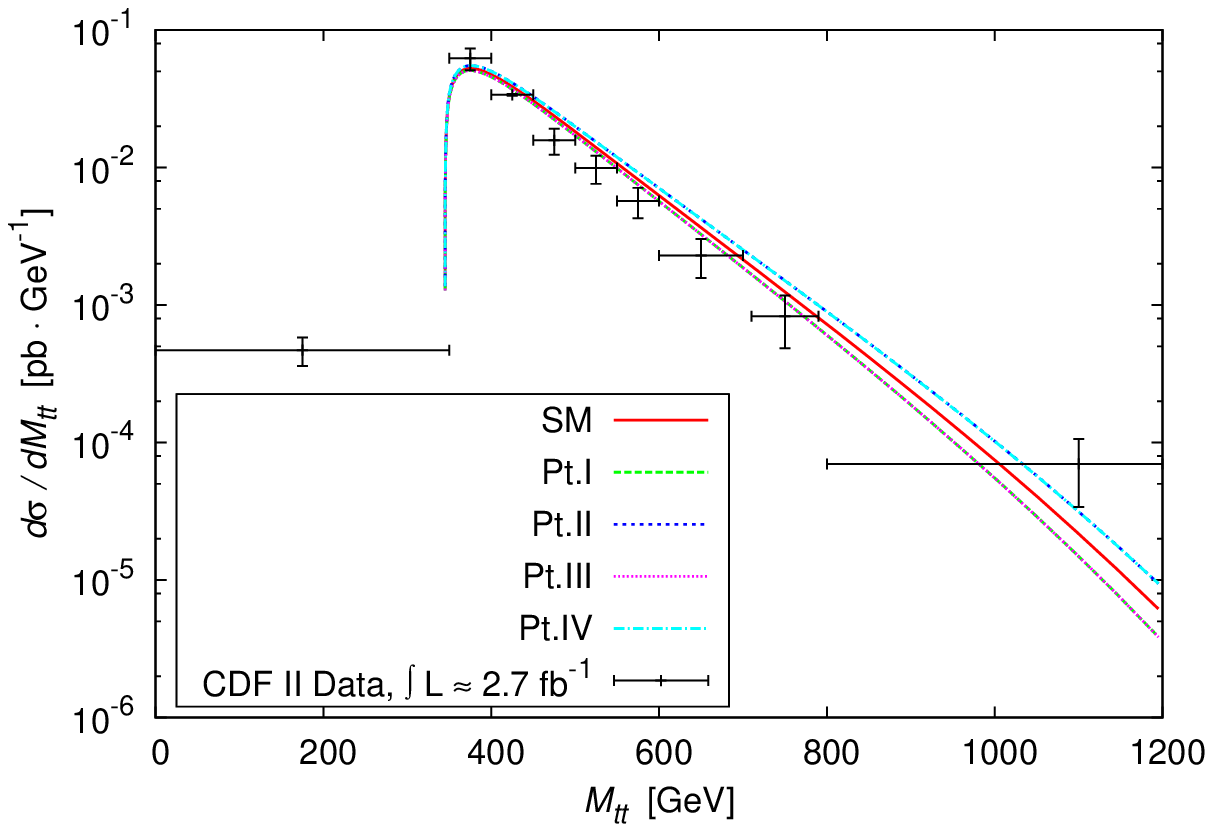}\\%
\includegraphics[width=7cm]{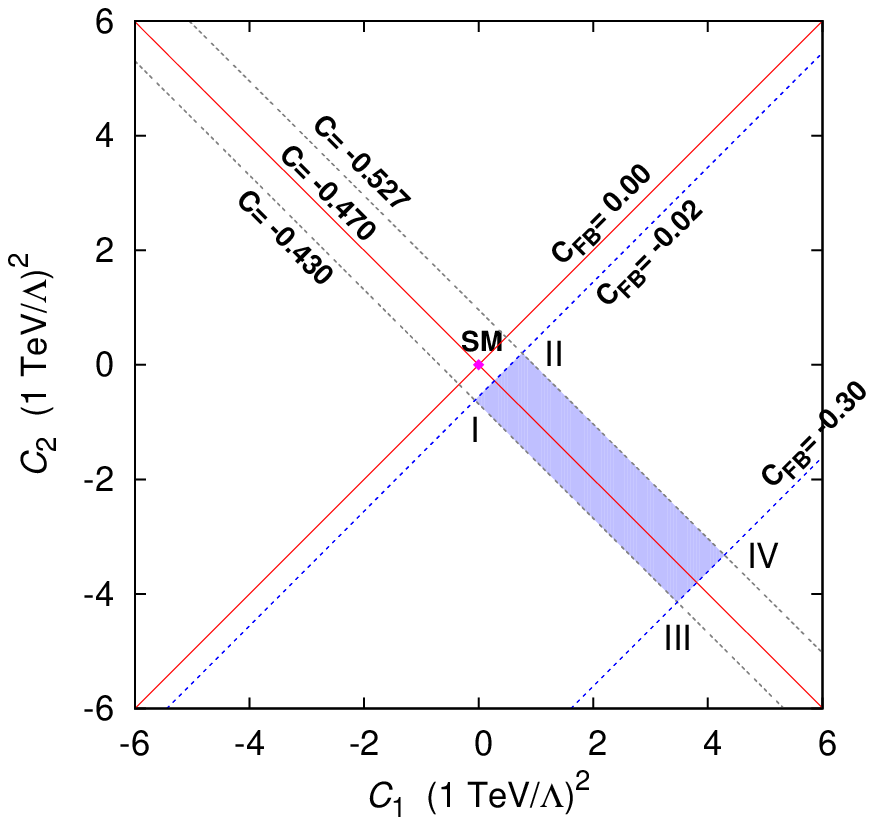}
\caption{\label{} 
(a) The $M_{t\bar{t}}$ distribution
for the points I,II,III, IV and the SM
and (b) the spin-spin correlations $C$ and $C_{FB}$.}
\label{fig:1bc}
\end{figure}

\medskip
{\bf\large 3.} It is straightforward to calculate the amplitude for 
\[
q (p_1) + \bar{q} (p_2) \rightarrow t (p_3) + \bar{t} (p_4) 
\]
using the above effective lagrangian and the SM. 
The squared amplitude summed (averaged) over the final (initial) 
spins and colors is given by
\begin{widetext}
\begin{eqnarray}
\overline{|{\cal M}|^2} 
\simeq  \frac{4\,g_s^4}{9\,\hat{s}^2} \left\{
2 m_t^2 \hat{s} \left[
1+\frac{\hat{s}}{2\Lambda^2}\,(C_1+C_2)
\right] s_{\hat\theta}^2 
+\frac{\hat{s}^2}{2}\left[ \left(1+\frac{\hat{s}}{2\Lambda^2}\,(C_1+C_2)\right)
(1+c_{\hat\theta}^2)
+\hat\beta_t\left(\frac{\hat{s}}{\Lambda^2}\,(C_1-C_2)\right)c_{\hat\theta}
\right]\right\}
\label{eq:ampsq}
\end{eqnarray}
\end{widetext}
where $\hat{s} = (p_1 + p_2)^2$, $\hat\beta_t^2=1-4m_t^2/\hat{s}$,
and $s_{\hat\theta}\equiv \sin\hat\theta$ and 
$c_{\hat\theta}\equiv \cos\hat\theta$ 
with $\hat{\theta}$ being the polar
angle between the incoming quark and the outgoing top quark in the 
$t\bar{t}$ rest frame. 
And the couplings are defined as:
$C_1 \equiv C_{8q}^{LL}+C_{8q}^{RR}$ and 
$C_2 \equiv C_{8q}^{LR}+C_{8q}^{RL}$ 
\footnote{Throughout this work, unless explicitly written, we are taking
$C^{AB}_{8q}=C^{AB}_{8u}=C^{AB}_{8d}$ assuming the 
$SU(2)_L \times SU(2)_R$ chiral symmetry. Under this assumption, 
the down-quark contribution to $\sigma_{t\bar{t}}$ and
$A_{\rm FB}$ is suppressed relative to the 
up-quark one by a factor more than $\sim 6$ at the Tevatron.}.
Since we have kept only up to the interference terms, there are 
no contributions from  the color-singlet operators with coupling 
$C_{1q}^{AB}$. The first (second)  term is for the production of top quarks with 
the same (different)  helicities. The term linear in $\cos\hat{\theta}$ could
generate the forward-backward asymmetry 
which is proportional to $\Delta C \equiv (C_1 - C_2)$.
Note that both light quark and top quark should have chiral couplings
to the new physics in order to generate $A_{\rm FB}$ at the tree level
(namely $\Delta C \neq 0$).  Therefore there could be parity violation in the 
four-quark operators with light quarks only, depending on the underlying physics.  
Since the SM neutral weak current effect is order of 
$g_2^2/m_Z^2$, it would be safe if $g_s^2 C / \Lambda^2 \lesssim 
g_2^2/m_Z^2$, which is not so difficult to achieve. 
This parity violation, if large, could be observed in the nonzero (anti)top 
spin polarization \cite{progress}. 

We convolute the parton level cross section with CTEQ6L parton distribution 
functions with factorization and renormalization scale equal to 
$m_t = 172.5$ GeV in order to get the cross sections and compared with the data
from the Tevatron. The QCD corrections are taken into account by the $K$ factor 
($K =1.3$) for the SM contributions in the total cross section. 
When we calculate the FB asymmetry, we use the LO results for the cross section.

In order to estimate the validity of our description 
based on effective lagrangian up to dimension 6 operators,  
we have to estimate the effects of the ignored dim-8 operators 
to the $t\bar{t}$ production at the Tevatron. 
For this purpose,  we calculate the amplitude squared from 
the effective lagrangian (2), see Appendix, and consider it as the rough estimate of 
dim-8 operators,  by comparing it to its interference term with the SM 
amplitude as well as to the SM amplitude squared.  
We note that, without considering all the possible dim-8 operators 
explicitly, the criterion adopted here for the validity of our description 
should  not be unambiguous, and intrinsically qualitative rather than 
quantitative. 

We find that the amplitude squared of dim-6 operators does not show 
simple dependence on the Wilson coefficients as the interference term.
Moreover, it includes additional contribution from the color-singlet operator 
which is proportional to $(C_{1q}^{AB})^2$ and enhanced by the factor $9/2$
compared to the color-octet contribution.  Since the color-singlet operators
do not interfere with the SM ones,  we set $C_{1q}^{AB} = 0$ 
and consider the validity region of our approach.

There are 3 inequalities we can consider:  
\begin{eqnarray}
\sigma_{\rm int} & < & r  \sigma_{\rm SM}  ,  \\
\sigma_{\rm NP} & < & r \sigma_{\rm int} ,  \\
\sigma_{\rm NP} & < & r^2 \sigma_{\rm SM} , 
\end{eqnarray}
where $\sigma_{\rm int}$ is the interference between the SM and the 
dim-6 operators, and $\sigma_{\rm NP}$ is the contributions solely
from the dim-6 operators.  The parameter $r$, a kind of fuzzy factor, 
measures the ratio of the subleading term to the leading $\sigma_{\rm SM}$ 
(SM contribution), and we take $r = 0.3 , 0.5 $ and 1.0 in the following.
We vary the couplings $C_{8q}^{AB}$ in the range between $-10$ and $+10$, 
and find the regions that satisfy the above inequalities.

The first inequality constrains the parameter space to the band around
$C_1 + C_2 = 0$. 
The 2nd relation implies that the contributions from dim-8 operators
are smaller than the dim-6 operators, and should be a good estimate 
in most cases. However, in our case, the interference $\sigma_{\rm int}$ 
is vanishing along the line $C_1 + C_2 = 0$, and we can not rely on 
the inequality, Eq.~(5). 
Therefore we also impose the 3rd inequality, Eq.~(6), which 
is relevant when $\sigma_{\rm NP} \propto ( C_1 + C_2 ) \approx 0$, 
and the constraint is the elliptical shape centered at the origin.  
Note that we impose the inequalities either (5) or (6) as the validity region, 
since we cannot apply the inequality (5) around the line $C_1 + C_2 =0$. 


In Fig.~\ref{fig:1a}, we show the allowed region in the $(C_1,C_2)$ plane 
that is consistent with the Tevatron data  at the $1 \sigma$ level:   
$\sigma_{t\bar{t}} = (7.50 \pm 0.48)$ pb  
and $A_{\rm FB} = (0.24 \pm 0.13 \pm 0.04)$.
The allowed region is around $0 \lesssim C_1 \lesssim 4$ and 
$-4 \lesssim  C_2 \lesssim + 0.5$. The negative sign of $C_2$ is preferred
at the 1 $\sigma$ level.   
The validity region described in the previous paragraph
is indicated in magenta, blue, black for $r=0.3, 0.5$ and 1, respectively. 
The straight line is from the inequality Eq.~(4), the circles and their mirror images are
from the inequality Eq.~(5), and the ellipse centered at the origin is from Eq.~(6).
Note that our effective lagrangian approach based on  dim-6 operators 
is valid in an ample parameter space.

In Fig.~\ref{fig:1bc} (a), we show the $\sqrt{\hat{s}}=M_{t\bar{t}}$ 
distributions  for four different values of $(C_1,C_2)$ corresponding to
the four corners of the allowed region in Fig.~\ref{fig:1a}, 
as well as the SM prediction.
The $M_{t\bar{t}}$ distribution depends only on $\sigma_{t\bar{t}}$, and there is no 
informations on the chiral structure 
because we have integrated out 
the angular dependence as well as  new heavy degrees of freedom 
that could modify $q\bar{q} \rightarrow t\bar{t}$.

\medskip
{\bf\large 4.} 
Another interesting observable which is sensitive to the chiral 
structure of new physics affecting $q\bar{q} \rightarrow t\bar{t}$ 
is the top quark spin-spin correlation \cite{Stelzer06}:
\beq
C = \frac{\sigma(t_L\bar{t}_L + t_R\bar{t}_R) - 
\sigma(t_L\bar{t}_R + t_R\bar{t}_L)}{\sigma(t_L\bar{t}_L + t_R\bar{t}_R) + 
\sigma(t_L\bar{t}_R + t_R\bar{t}_L)} \,.
\eeq
This quantity depends on the spin quantization axis, and 
there are three different axes that are mostly used: the beam axis, 
the top direction (helicity basis), and the off-diagonal axis. 
At leading order,
the SM prediction is $C= -0.471$ 
for the helicity basis which we choose in this work. 
It is known that NLO correction to $C$ is rather large, shifting 
$C$ to $-0.352$~\cite{Stelzer06}, but it is beyond the scope of this work to pursue
this issue further here.  

Recently the top spin-spin  correlation was measured by both 
CDF and D0 Collaborations at the Tevatron \cite{heinz}. 
Using the off-diagonal axis as the quantization axis 
in order to maximize the spin-spin correlation, they found 
$C= 0.32^{+0.55}_{-0.78}$ (CDF dilepton mode), 
$C=0.60 \pm 0.50({\rm stat}) \pm 0.16({\rm syst})$
(CDF lepton + jets mode), and 
$C=-0.17^{+0.64}_{-0.53}$ (D0 dilepton mode).
The data are consistent with the SM prediction $C=0.78$ (for the 
off-diagonal basis), albeit rather large uncertainties. 
It would be interesting if the SM prediction is confirmed or not
with more accumulated data in the future.

Since new physics must have chiral couplings both to light quarks and 
top quark, the spin-spin correlation defined above will be affected.
From Eqs.~(3) and (7),  it is clear this correlation is sensitive to $(C_1 + C_2)$, 
since the linear term in $\cos\hat{\theta}$ does not contribute to
the correlation $C$ after integration over $\cos\hat{\theta}$.
On the other hand, if one considers the forward and the backward regions 
separately,  the spin-spin correlation would depend on $( C_1 - C_2 )$
and will be closely correlated with $A_{\rm FB}$. 
Therefore we propose a new spin-spin FB asymmetry $C_{FB}$  defined as 
\begin{equation}
C_{FB} \equiv  C (\cos\theta \geq 0) -  C (\cos\theta \leq 0)  ,
\end{equation}
where $C(\cos\theta \geq 0 (\leq 0))$ implies the cross sections in the 
numerator of Eq.~(7) are obtained for the forward (backward) region: 
$\cos\theta \geq 0 (\leq 0)$.
This quantity can be measured by dividing the $t\bar{t}$ sample into 
the forward top and the backward top events. 
In Fig.~\ref{fig:1bc} (b),  we show the contour plots for the $C$ and $C_{FB}$ 
in the $(C_1 , C_2 )$ plane along with the SM prediction at LO.
Note that there is a clear correlation between 
$C_{FB}$ and $A_{FB}$ in Fig.~\ref{fig:1a}, which must be observed in the 
future measurements if the $A_{\rm FB}$ anomaly is real and  
a new particle is too heavy to be produced at the Tevatron.  


\medskip
{\bf\large 5.}
So far, we considered dim-6 four-quark operators that could
affect the $t\bar{t}$ productions at the Tevatron, and found 
the necessary conditions for accommodating $A_{\rm FB}$.
Now we study specific new physics that could generate the 
relevant dim-6 operators with corresponding Wilson coefficients.
It is impossible to exhaust all the possibilities, and we consider
the following interactions of quarks with spin-1 
flavor-conserving (FC) color-octet $V^a_8$ vectors,
spin-1 flavor-violating (FV) color-singlet $\tilde{V}_1$ and color-octet 
$\tilde{V}^a_8$ vectors, and spin-0 FV color-singlet $\tilde{S_1}$
and color-octet $\tilde{S}^a_8$ scalars ($A=L,R$):
\beqr
&& \mathcal{L}_{\rm int} =  
g_s V_8^{a\mu}\sum_{A}\left[g_{8q}^A(\bar{q}_A\gamma_\mu T^a q_A) 
+ g_{8t}^A(\bar{t}_A\gamma_\mu T^a t_A) \right]    
\nonumber  \\
&& 
+ g_s\big[\tilde{V}_1^\mu \sum_{A}\tilde{g}_{1q}^A(\bar{t}_A\gamma_\mu q_A)
+ \tilde{V}_8^{a\mu} \sum_{A}\tilde{g}_{8q}^A(\bar{t}_A\gamma_\mu T^a q_A) 
+ \textrm{h.c.} \big]    \nonumber \\
&& + g_s\big[\tilde{S}_1 \sum_{A}\tilde{\eta}_{1q}^A(\bar{t} A q)
+ \tilde{S}_8^{a} \sum_{A}\tilde{\eta}_{8q}^A(\bar{t} A T^a q) + \textrm{h.c.} \big] ,
\eeqr 
%
where $q$ denotes light quarks (either $u$ or $d$ depending on the models).
This interaction lagrangian encompasses many models beyond the SM, and 
makes a good starting point to study the underlying mechanism for 
the effective lagrangian discussed earlier.
If the spin-1 particle has both the FC and FV interactions,  
we set $V_{8}^\mu =\tilde{V}_{8}^\mu$.
The axigluon model \cite{axigluon} corresponds to flavor universal chiral 
couplings: $g_{8q}^L = g_{8t}^L = - g_{8q}^R = - g_{8t}^R = 1$. 
The model in Ref.~\cite{Jung:2009jz} has a new gauge boson $Z^{'}$ with dominant
couplings to the $u$ and $t$ quarks,  which in our language corresponds to 
$\tilde{V}_1 = Z^{'}$ and $g_s \tilde{g}_{1q}^R = g_X$ ignoring the FC coupling.
The model in Ref.~\cite{Cheung:2009ch} has a new charged gauge boson 
$W^{\pm '}$, which is obtained in our case by $\tilde{V}_1 = W^{'}$  
and $g_s \tilde{g}_{1q}^A = g^{'} g_A$ for $A=L$ or $R$, respectively. 
In some RS scenarios, one can have large flavor mixing in the right-handed 
quark sector \cite{Agashe:2006hk}: 
$g_{8q}^L = g_{8q}^R = g_{8b}^R \simeq -0.2$, 
$g_{8t}^L = g_{8b}^L \simeq (1 \sim 2.8)$, $g_{8t}^R \simeq (1.5 \sim 5)$, 
$\tilde{g}_{8q}^L \simeq V_{tq}$ and $\tilde{g}_{8q}^R \simeq 1$.

After integrating out the heavy vector and scalar fields, we obtain 
the Wilson coefficients as follows:
\beqr
\frac{C^{LL}_{8q}}{\Lambda^2} &=& -\frac{1}{m_V^2} g_{8q}^L g_{8t}^{L} -
 \frac{1}{m^2_{\tilde{V}}}\left[2 |\tilde{g}_{1q}^L |^2
  - \frac{1}{N_c}|\tilde{g}_{8q}^L |^2 \right]\,,  \nonumber   \\
\frac{C^{RR}_{8q}}{\Lambda^2} &=& -\frac{1}{m_V^2} g_{8q}^R g_{8t}^{R}
 - \frac{1}{m^2_{\tilde{V}}}\left[2 |\tilde{g}_{1q}^R |^2 
 - \frac{1}{N_c}|\tilde{g}_{8q}^R |^2 \right] \,,  \label{eq:wcs1} \\
\frac{C^{LR}_{8q}}{\Lambda^2} &=& -\frac{1}{m_V^2} g_{8q}^L g_{8t}^{R} 
- \frac{1}{m^2_{\tilde{S}}}\left[ |\tilde{\eta}_{1q}^L|^2 
- \frac{1}{2N_c}|\tilde{\eta}_{8q}^L|^2 \right]\,, \nonumber  \\
\frac{C^{RL}_{8q}}{\Lambda^2} &=& -\frac{1}{m_V^2} g_{8q}^R g_{8t}^{L} 
- \frac{1}{m^2_{\tilde{S}}}\left[ |\tilde{\eta}_{1q}^R|^2 
- \frac{1}{2N_c}|\tilde{\eta}_{8q}^R|^2 \right]\,, 
\nonumber  
\eeqr
where $m_V(m_{\tilde{V}})$ and $m_{\tilde{S}}$ denote the masses of vectors 
$V_8\, (\tilde{V_i})$ and scalars $\tilde{S_i}$ $(i=1,8)$, respectively.

Another interesting possibility is minimal flavor violating interactions of 
color-triplet $S_k^\gamma$ with mass $m_{S_3}$ and 
color-sextet scalars $S_{ij}^{\alpha\beta}$ with mass
$m_{S_6}$ with the SM quarks \cite{Arnold:2009ay}. Here 
$\alpha, \beta, \gamma$ and $i,j,k$ are color and flavor indices, respectively. 
For example, if we consider the following interactions (Model V and VI in 
Ref.~\cite{Arnold:2009ay}), 
\beq
{\cal L} = g_s \Big[ \frac{\eta_3}{2} \epsilon_{\alpha\beta\gamma} \epsilon^{ijk} 
u_{iR}^\alpha u_{jR}^\beta S_k^\gamma 
+  \eta_6 u_{iR}^\alpha u_{jR}^\beta S_{ij}^{\alpha\beta} + h.c. \Big]
\eeq 
the $u-$channel exchange of new scalars can contribute to 
$u \bar{u} \rightarrow t \bar{t}$, resulting in 
\footnote{ 
$C_{1q}^{RR}$ is also induced by color-triplet and 
sextet scalars,  but is not shown, since it is irrelevant here.
} 
\beq 
\label{eq:wcs2}
\frac{C_{8u}^{RR}}{\Lambda^2}   =  - \frac{| \eta_3 |^2}{ m_{S_3}^2} +  
\frac{2 |\eta_6|^2}{m_{S_6}^2} \ .
\eeq 
Since these new scalars couple only to the right-handed up-type quarks, 
constraints on the couplings $\eta_3$ and $\eta_8$ from flavor physics
are rather weak, and one can accommodate the observed $A_{\rm FB}$ easily.

\begin{figure}
\hspace{-1.5cm}
\includegraphics[width=10cm]{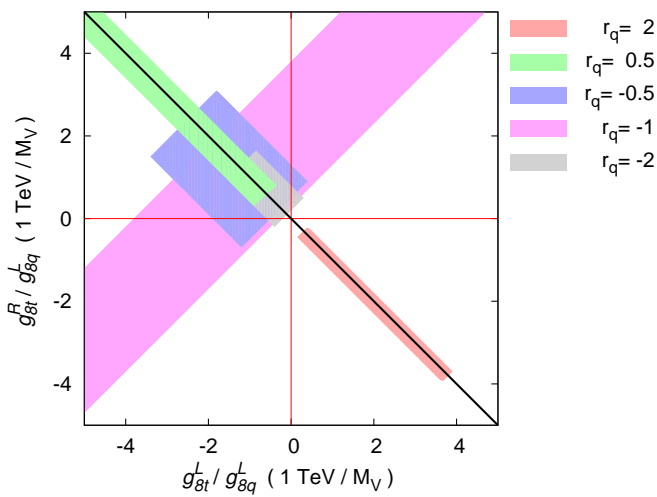}
\caption{\label{} The 1-$\sigma$ favored region of the couplings for
the exchanges of the spin-1 FC color-octet vector bosons $V_8$. }
\label{fig:2}
\end{figure}

In Table~\ref{tab:newparticles}, we show the new particle exchanges  
under consideration and the couplings $C_1, C_2$ induced by them.
We found that the four types of exchanges of $V_8$, $\tilde{V}_8$,
$\tilde{S}_1$, and $S_{13}^{\alpha\beta}$ could give rise to the large positive 
$A_{\rm FB}$ at the 1-$\sigma$ level.
Among them, the following three types of couplings 
are more or less fixed as:
\begin{eqnarray}
\tilde{V}_8    &:&  
  \frac{1}{N_c}\,\left(\frac{1\,{\rm TeV}}{m_{\tilde{V}}}\right)^2\, 
\left( |\tilde{g}_{8q}^L |^2 
  +|\tilde{g}_{8q}^R |^2  \right) \simeq 0.76 \,, \nonumber \\
\tilde{S}_1    &:&  
  \left(\frac{1\,{\rm TeV}}{m_{\tilde{S}}}\right)^2\, \left( |\tilde{\eta}_{1q}^L |^2 
  +|\tilde{\eta}_{1q}^R |^2  \right) \simeq 0.62 \,, \nonumber \\
{S}_{13}^{\alpha\beta}    &:&  
  2\,\left(\frac{1\,{\rm TeV}}{m_{S_6}}\right)^2\, |\eta_6|^2 \simeq 0.76\ .
\end{eqnarray}
For the case of $V_8$, Fig.~\ref{fig:2} shows  the 1-$\sigma$ favored regions 
of the couplings in the $(g^{L}_{8t},g^{R}_{8t})$ plane for several choices 
of $r_q \equiv g^{R}_{8q}/g^{L}_{8q}$ taking $g^{L}_{8q}=1$.
We observe that the regions lie along the direction of $g^{L}_{8t}=-g^{R}_{8t}$ and
$g^{L}_{8t}/g^{L}_{8q}$ tends to be positive (negative) for $r_q>1$ $(r_q<1)$.
Note that $r_q=1$ is not allowed because $\Delta C=0$ and
the constraint from $\sigma_{t\bar{t}}$ disappears when $r_q=-1$, 
since $C_1+C_2=0$.  
It would be interesting to search for new vector or 
scalar particles that satisfy the above conditions at LHC.

A caution is in order. The second row with $\tilde{V}_1$ in Table I is 
shown to be disfavored, which seems to be contradictory to the results
of Refs.~\cite{Jung:2009jz,Cheung:2009ch} where new particles were 
assumed to be light. 
However this is not the case, since we 
took a heavy new particle limit ( far below the new particle
mass scale, $\hat{s}, \hat{t} \ll m_{\tilde{V}}^2$) in order to derive 
$C_{8q}^{AB}$'s, though we
began with the full amplitudes using the above interaction 
lagrangian which are the same as the amplitudes in  
Refs.~\cite{Jung:2009jz,Cheung:2009ch}.
Therefore our results can not be in conflict with
the claims of Refs.~\cite{Jung:2009jz,Cheung:2009ch}.
Rather, our results simply imply that a color-singlet spin-1 object cannot 
accommodate the $A_{\rm FB}$ if it is too heavy to be directly 
produced at the Tevatron. 
In other words, one has to consider a light $\tilde{V}_1$ scenario, where  
$\tilde{V}_1$ has to be light enough ($\sim O(100)$ GeV) as in 
Refs.~\cite{Jung:2009jz,Cheung:2009ch}, 
and then our effective lagrangian approach is not applicable 
for such a light $\tilde{V}_1$.

On the other hand, a new particle with different quantum numbers 
(color, spin and flavor mixings) can affect the $A_{\rm FB}$, even if 
it is too heavy to be produced directly at the Tevatron. 
As an example, one can consider the color-sextet case in order to  
make clear our point.  In this case, only $C_{8u}^{RR}$ (and $C_{1u}^{RR}$ 
which is irrelevant in the interference) is nonzero, and we have
$C_2 =0$. Then Fig.~\ref{fig:1a} indicates that a very heavy color-sextet can
enhance $A_{\rm FB}$ roughly up to $\sim 12 \%$ at the $1\sigma$ 
deviation in the total cross section $\sigma_{t\bar{t}}$ to the larger value.
For the color antitriplet case, the signs of $C_{8u}^{RR}$ and $C_1$ are 
opposite to the color sextet case [ see Eq.~(12) ], and the resulting 
$A_{\rm FB}$ is less than the color-sextet case, with affecting 
$\sigma_{t\bar{t}}$ in the lower value.  These tendency can be shown 
to be consistent with the calculations based on the full amplitudes
\cite{Jung:2009jz,Cheung:2009ch,progress}.
In order to increase $A_{\rm FB}$ more, there would be a strong tension
with the total cross section $\sigma_{t\bar{t}}$, if a new physics scale is
beyond the reach of Tevatron. However this tension might be avoided 
(or lessened) if the new particle is light enough, and we should not 
rely on the effective lagrangian approach for this case.
One can understand other entries of Table~I in the similar fashion.

For more quantitative discussions,  we have to study the full amplitude and 
compare the results with our effective lagrangian approach, and investigate 
the validity region of the effective lagrangian approach model by model.
This is clearly beyond the scope of the present work, and the detailed 
study will be presented in the future work \cite{progress}.

\begin{table}[t]
\caption{\label{tab:newparticles}
{\it
New particle exchanges and the signs of induced couplings $C_1$ and $C_2$}
}
\begin{center}
\begin{tabular}{l|c|c|c|c}
\hline\hline
& & & & \\[-0.3cm]
New particles & couplings & $C_1$ & $C_2$ & 1 $\sigma$ favor \\[0.1cm]
\hline\hline
& & & & \\[-0.3cm]
$V_8$ (spin-1 FC octet) & $g^{L,R}_{\,8q,8t}$ & indef. & indef. & $\surd$ \\[0.1cm]
$\tilde{V}_1$ (spin-1 FV singlet) & $\tilde{g}^{L,R}_{1q}$ & $-$ & $0$ & $\times$ \\[0.1cm]
$\tilde{V}_8$ (spin-1 FV octet) & $\tilde{g}^{L,R}_{8q}$ & $+$ & $0$ & $\surd$ \\[0.1cm]
\hline
& & & & \\[-0.2cm]
$\tilde{S}_1$ ~~(spin-0 FV singlet) & $\tilde\eta^{L,R}_{1q}$ & $0$ & $-$ & $\surd$ \\[0.1cm]
$\tilde{S}_8$ ~~(spin-0 FV octet) & $\tilde\eta^{L,R}_{8q}$ & $0$ & $+$ & $\times$ \\[0.1cm]
$S_2^\alpha$ ~\,(spin-0 FV triplet) & $\eta_{3}$ & $-$ & $0$ & $\times$ \\[0.1cm]
$S_{13}^{\alpha\beta}$ \,(spin-0 FV sextet) & $\eta_{6}$ & $+$ & $0$ & $\surd$ \\[0.1cm]
\hline\hline
\end{tabular}
\end{center}
\end{table}

\medskip
{\bf\large 6.}
In this letter,  we performed a model independent study
of $t\bar{t}$ productions at the Tevatron 
using dimension-6 contact interactions relevant to 
$q\bar{q}\rightarrow t \bar{t}$. 
We derived conditions for the couplings
of four-quark operators that could generate the FB asymmetry 
observed at the Tevatron [Fig.~\ref{fig:1a}]. 
Then we considered the $s-$, $t-$ and $u-$channel exchanges of 
spin-0 and spin-1 particles whose color quantum number is 
either singlet, octet, triplet or sextet.   
Our results in Eqs.~(\ref{eq:wcs1}), (\ref{eq:wcs2}), and (13) 
and Fig.~\ref{fig:2}
encode 
the necessary conditions for the underlying new physics 
in a compact and an effective way, when those new particles
are too heavy to be produced at the Tevatron but still affect $A_{\rm FB}$.  
If these new particles could be produced directly at the Tevatron or 
at the LHC,  we cannot use the effective lagrangian any more. 
We have to study specific models case by case, and anticipate 
rich phenomenology at colliders as well as at low energy. 
Detailed study of these issues lies beyond the scope of this letter, 
and will be discussed in the future publications \cite{progress}. 


\medskip

\begin{acknowledgements}
We are grateful to S. Choi, D.H. Kim, S.B. Kim and I. Yu for 
useful communications, and S.C.Park for collaboration in 
the initial stage of this work. 
Part of works by PK were done at Aspen 
Center for Physics, and Galileo Galilei Institute for Theoretical Physics.   
The work of Nam was supported in part by Basic Science Research Program 
through the National Research Foundation of Korea (NRF) funded by 
the Ministry of Education, Science and Technology (2009-0086961).
\end{acknowledgements}

\medskip

\setcounter{equation}{0}
\section*{Appendix}
\begin{appendix}

In  this Appendix, we present the explicit form of
the amplitude squared from the effective lagrangian (2), which is used to
calculate $\sigma_{\rm NP}$, as follows:
\begin{widetext}
\begin{eqnarray}
\overline{|{\cal M}_{\rm NP}|^2}
&=&\frac{4 g_s^4}{9 \hat{s}^2} \frac{\hat{s}^2}{4 \Lambda^4}
\left\{ 
 \left[  9 \left( (C_{1q}^{LL})^2+(C_{1q}^{RR})^2 \right)
            + 2 \left( (C_{8q}^{LL})^2+(C_{8q}^{RR})^2 \right) \right]
\left(\hat{u}-m_t^2 \right)^2 
\right. 
\nonumber \\[0.2cm]
&& \hspace{1.2cm}
+ \left[  9 \left( (C_{1q}^{RL})^2+(C_{1q}^{LR})^2 \right)
            + 2 \left( (C_{8q}^{RL})^2+(C_{8q}^{LR})^2 \right) \right]
\left(\hat{t}-m_t^2 \right)^2 \nonumber \\[0.2cm]
&& \hspace{1.15cm}
+ \left.
  \left[  9 \left(  C_{1q}^{LL} C_{1q}^{LR}+C_{1q}^{RR} C_{1q}^{RL} \right)
     + 2 \left(  C_{8q}^{LL}  C_{8q}^{LR}+C_{8q}^{RR}  C_{8q}^{RL} \right) 
     \right] \left(2 \hat{s} m_t^2 \right) 
\right\} \,,
\end{eqnarray}
where $\hat{u}-m_t^2=-\hat{s}\,(1+\hat\beta_t c_{\hat\theta})/2$
and $\hat{t}-m_t^2=-\hat{s}\,(1-\hat\beta_t c_{\hat\theta})/2$.
\end{widetext}
\end{appendix}



%

\end{document}